# Millimeter Wave and Terahertz Synthetic Aperture Radar for Locating Metallic Scatterers Embedded in Scattering Media

Jonathan T. Richard,[1] Henry O. Everitt [2]

[1] IERUS Technologies, Huntsville, AL 35805, USA.

[2] Charles M. Bowden Research Center, Army Aviation & Missile RD&E Center, Redstone Arsenal, AL 35898, USA.

**Abstract**: A rail-mounted synthetic aperture radar has been constructed to operate at W-band (75 - 110 GHz) and a THz band (325 - 500 GHz) in order to ascertain its ability to locate isolated small, visually obscured metallic scatterers embedded in highly scattering dielectric hosts that are either semi-transparent or opaque. A "top view" 2D algorithm was used to reconstruct scenes from the acquired data, locating metallic scatterers at W-band with high range and cross-range resolution of 4.3 and 2 mm, respectively, and with improved range resolution of 0.86 mm at the THz band. Millimeter-sized metallic scatterers were easily located when embedded in semi-transparent, highly scattering target hosts of Styrofoam and waxy packing foam but were more difficult to locate when embedded in relatively opaque, highly scattering Celotex panels. Although the THz band provided the expected greater spatial resolution, it required the target to be moved closer to the rail and had a more limited field of view that prevented some targets from being identified. Techniques for improving the signal to noise ratio are discussed. This work establishes a path for developing techniques to render a complete 3D reconstruction of a scene.

# 1. Introduction

Synthetic Aperture Radar (SAR) is a well-established, widely fielded technique that uses either a moving platform or a moving target to reconstruct images with higher spatial resolution than the diffraction-limited imagery possible with a stationary aperture and target [1][2][3][4][5]. Most deployed SARs operate at conventional microwave frequencies because of the high powers and sufficient bandwidth available to reconstruct large scenes (>>1 km$^2$) over long distances (>1 km) with good resolution (<10 m). Considerably less explored is the application of compact SAR for short range, high resolution applications such as non-destructive test of composite structures, location of scatterers in opaque media, and non-contact surveying of compact, complex objects such as those fabricated by modern 3D printing techniques. For such high resolution (< 1 cm) applications, high frequency SAR must be developed and deployed on platforms that are stable to within a fraction of a wavelength or are extremely well characterized in real time so that the location of the transceiver can reliably be known and accurate reconstructions made. Some progress has been made to develop millimeter wave and even terahertz frequency SAR, including some impressive proof-of-concept demonstrations, but much work remains for these high frequency radars to become practical [6][7][8][9]. The clear technological challenges in source power, component fragility and cost, and platform stability are further exasperated by the increasing atmospheric attenuation with increasing frequency, suggesting that the application of such radars will exploit its inherently shorter range and higher resolution.

Here we demonstrate millimeter wave (MMW) and terahertz (THz) frequency SAR reconstructions using a transceiver mounted on a rail in order to resolve millimeter-sized metallic scatterers embedded in highly scattering host panels composed of semi-transparent foams or an opaque thermoset plastic. Two different frequency bands were compared to quantify the tradeoff of increased resolution and decreased penetration. Two separate 2D SAR reconstruction algorithms were considered with the ultimate goal of achieving a 3D SAR reconstruction of the scene. Side-looking SAR provided the best 2D reconstructions in this study. Several target hosts, including extruded polystyrene foam (Styrofoam) and polyisocyanurate insulation boards (Celotex), were used to characterize the transmissibility of the medium for the frequency bands of interest. Styrofoam was transparent in both bands while the opacity of Celotex reduced the signal-to-noise ratio of the embedded metallic scatterers and made object identification more difficult. Moreover, clutter from both hosts produced false positive returns that had to be discriminated from the metallic scatterers. Our objective was to establish and assess the necessary tools in 2D to achieve full 3D rail SAR reconstruction at high frequencies so that small scatterers may be located in absorptive, scattering hosts. Our work illustrates some of the opportunities and challenges these 3D reconstructions will face.

## 2. Experiment

The experimental system shown in Fig. 1(a,b) used an Agilent N5222A vector network analyzer (VNA) with MMW or THz transceiver front ends, a linear stage on which the transceiver was scanned, a tripod on which objects were mounted, and a computer to control the hardware and acquire the data. A Virginia Diodes (VDI) WR10 and WR2.2 transceivers transmitted and received W-band (75 - 110 GHz) and THz (325 - 500 GHz) signals, and radiation was coupled through a standard gain horn antenna. The VDI W band transceiver includes the necessary transmitter, receiver, and couplers within a single box using a single horn antenna, while the VDI THz transceiver consists of a separate VDI transmitter, two VDI receivers, and two couplers on a common breadboard with a single horn antenna. (The first coupler connects the transmitter to one receiver to establish a reference signal, and the second coupler connects to the second receiver to acquire the round trip-delayed signal which is compared to the reference signal.) Calibration of the VNA removed undesired reflections and standing waves that were consistently seen by the radar. The VNA was swept in frequency over the entire W-band or THz-Band with an intermediate frequency (IF) bandwidth of only 100 Hz to minimize noise.

Cross-range and range data were obtained by a lateral translation of the transceiver and a linear frequency sweep, respectively. The transceiver was mounted on a linear stage and moved 1 m in 500 discrete steps of length $\delta x = 2$ mm. At each location the transceiver emitted a linear frequency sweep, and the VNA recorded the received complex valued (real and imaginary IQ data) signal and digitized it into 4096 frequency bins. A Hamming window [19] was subsequently applied to the Fourier transform of the data to smooth it and lower sidelobes caused by spectral leakage. Finally, a Matlab algorithm computed an inverse fast Fourier transform (IFFT) of the data to convert the frequency domain data to the range domain, whose range resolution is

$$\Delta r = \frac{c}{2B} \qquad (1)$$

for frequency bandwidth $B$ in Hz and the speed of light $c$ in m/s. For this experiment, the W-band range resolution achieved with a bandwidth of $B = 35$ GHz is $\Delta r = 4.3$ mm, while the range resolution for the THz bandwidth $B = 175$ GHz was 0.86 mm. This provides an unambiguous range of $4096\Delta r = 17.6$ m for the W-band radar and 3.5 m for the THz radar. When completed, the collected 2D data fills an array of 500 x 4096 elements from which the scene may be reconstructed.

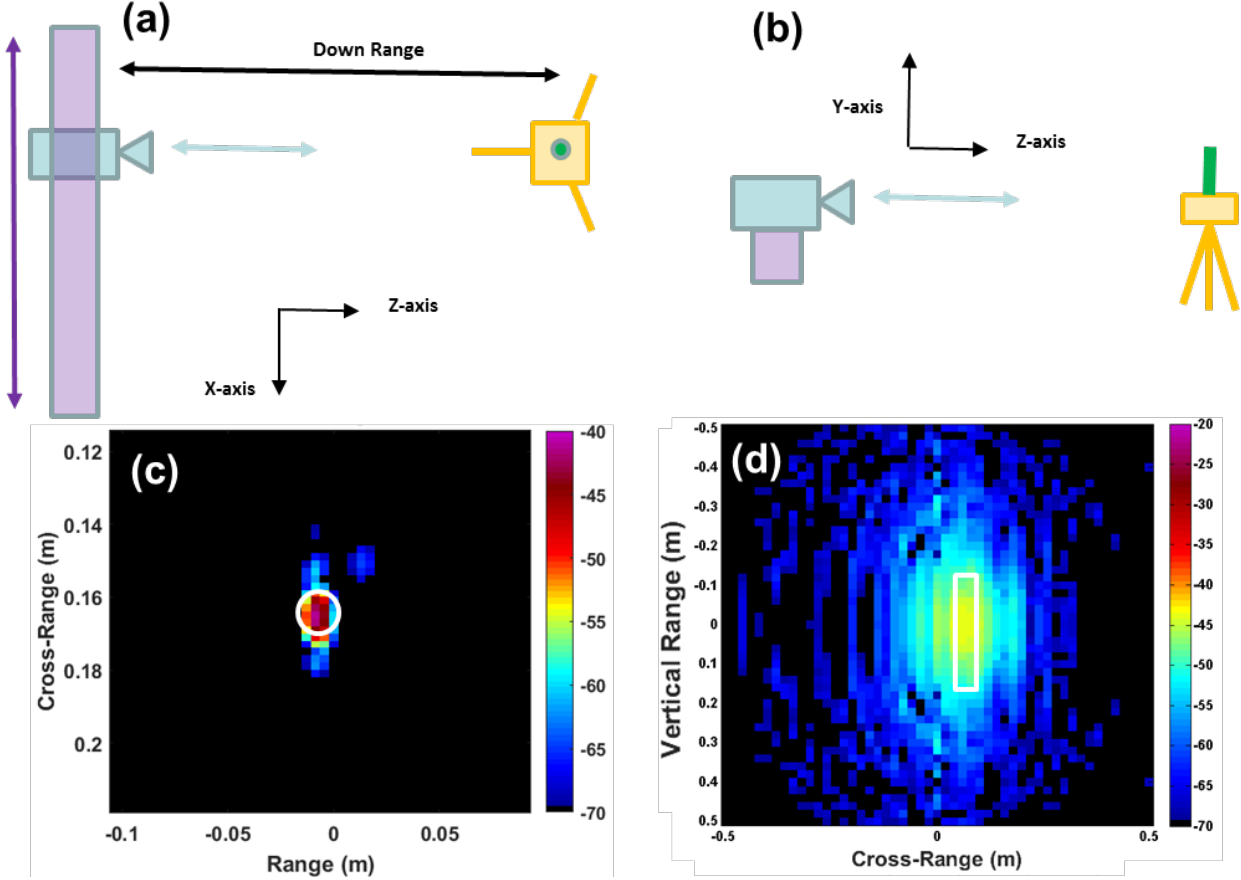

Fig. 1. (a) Top down view of the SAR setup. The range domain corresponds to the Z axis and the cross range domain corresponds to the X axis. (b) Side view of the SAR setup. The range domain corresponds to the Z axis and the cross vertical domain corresponds to the Y axis. Stripmap SAR reconstructions using the (c) Carrara X-Z and (d) Gorham Y-Z algorithms, with strengths presented in dB.

Although the ultimate goal is to be able to generate 3D SAR reconstructions, it is valuable first to consider various 2D SAR acquisition and reconstruction methodologies. The configurations shown in Fig. 1 provide for 2D scene reconstructions in the X-Z or Y-Z planes, respectively. For either configuration, rail SAR, also known as stripmap SAR, uses an antenna that emits a diverging beam, while spotlight SAR adjusts the beam to remain focused on a desired area as the antenna moves in location. The well-known algorithm by Carrara et al. is routinely used to reconstruct scenes in the X-Z configuration (Fig. 1(c)) and will be described more below [1]. For the side-looking X-Y configuration at a particular range along the Z axis, the Gorham and Rigling SAR algorithm may be used to generate 2D reconstructions (Fig. 1(d)) [11]. A 3D SAR reconstruction could be generated by utilizing a combination of Carrara and Gorham algorithms, for example, or a complete 3D SAR algorithm could be implemented [12][13][14][15][16][17][18]. In the absence of a 3D algorithm, a series of 2D scans with an increasingly tilted target or transceiver can remove the location ambiguity of scatterers distributed in the third dimension (e.g. for the Carrara algorithm, separating scatterers at the same X-Z location but separated in Y) and partially reconstruct the scene as scatterers project into one of the other two dimensions (e.g. in the X or Z direction).

## 3. Results and Discussion

Here we used stripmap SAR and the Carrara algorithm to produce top-down, X-Z 2D reconstructions of a variety of targets. The insights about the sensitivity and accuracy of the reconstructions provided by these analyses will be described below. For all of the datasets, the radar is looking at the target from the left. Initially a single metallic rod in free space was used to verify that the technique was working, and our procedure is detailed below for that simple target. Then more complex targets were investigated in order to estimate the spatial resolution and penetration depth in increasingly challenging configurations in increasingly absorptive and scattering hosts. These complex targets were composed of a variety of Styrofoam and Celotex host panels embedded with a variety of aluminum rods, rectangular brackets, staples, and strips of wire that approximate perfect electrical conductors (PECs).

## A. Single aluminum rod

Our first measurement was of an isolated cylindrical 1.3 cm diameter aluminum rod to illustrate the how the rail SAR technique is used to acquire and process a 2D data set and to ascertain how accurately the location of the rod could be quantified. The rod was placed at $z_0 = 2.3$ meters down range and $x_0 = 0.17$ meters in cross range from the center of the rail, as illustrated in Fig. 2(a). The VDI transceiver was stepped 500 times with 2 mm increments along the rail, and at each location the VDI transceiver performed a frequency sweep to collect 4096 points of IQ data spanning 75 - 110 GHz using a 100 Hz IF bandwidth. This produced a two dimensional data set composed of IQ data as a function of location (x) and frequency (related to z by an IFFT), plotted in Fig. 2(b).

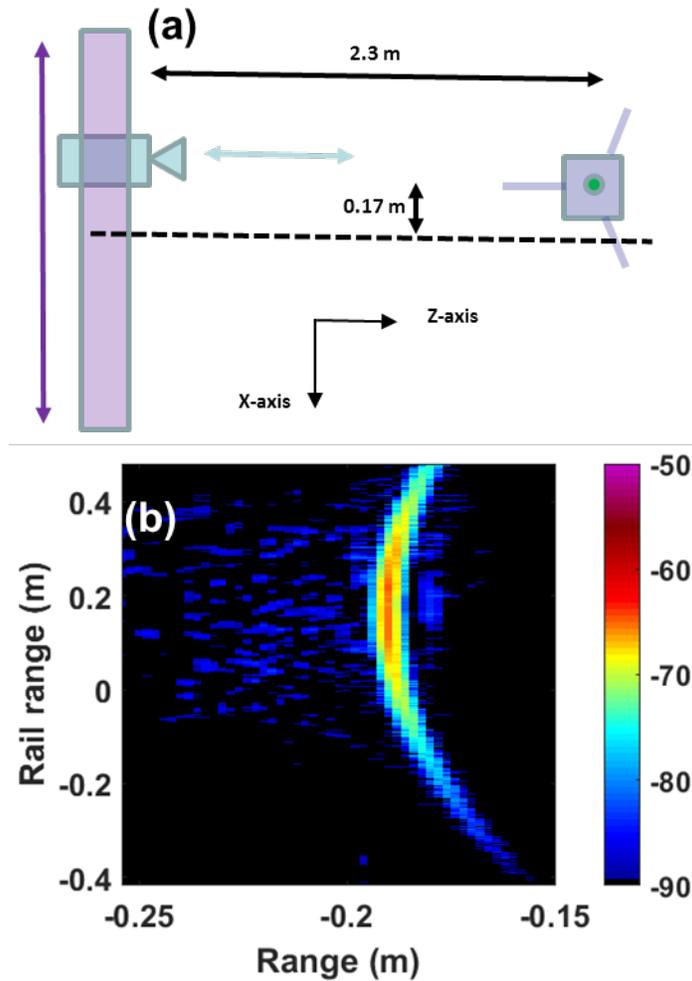

Fig. 2. (a) Top down view of the SAR setup. The rod is 2.3 meters downrange on the Z axis and 0.17 m cross-range on the X-axis. (b) Plot of a W-band 2D SAR data collected from a single rod, plotted in dB as a function of range and cross-range rail distances.

In stripmap SAR the distance of the rod from the transceiver changes as the transceiver scans, reaching a minimum when the transceiver is at $(x_0, 0)$ and increasing as the transceiver moves away on either side as $r = \sqrt{z_0^2 + (x - x_0)^2}$. The 2D unprocessed data plot in Fig. 2(b) clearly indicates the presence of this single scatterer through the boomerang shaped return measured, and the apex of the curve indicates the location of rod at $(x_0, z_0)$. To account for this boomerang effect, we adopted the Carrara algorithm to reconstruct the scene in the X-Z plane and measure the location ($x_0$, $z_0$) of the rod, as shown in Fig. 1(c) [1]. A range migration algorithm (RMA) interprets the range curvature produced by the scanning transceiver and accurately locates the target, following the procedure in [1]. First, an FFT operation, preceded and followed by FFT shift operations on the data, is performed along the cross range step dimension. Next, a 2D phase compensation matched filter $e^{jF_{mf}}$ is applied to this 2D data array to transform and correct for this curvature, where $F_{mf} = r_d(\sqrt{k_R^2 - k_x^2} - k_R)$. Here, $r_d$ is the range to the center of the scene, $k_R = \frac{4\pi}{\lambda}$ is the frequency wave number, and $k_x$ is the azimuthal wave number, which is represented as a 500 element linear array between $-\frac{\pi}{\delta x}$ and $+\frac{\pi}{\delta x}$ with step size $\delta x$ = 2 mm. Next, a Stolt interpolation [1] is applied to map the non-planar wavefront to the discretized 2D array across the step dimension of the data by using a 1D spline interpolation from $K_Y$ values to rounded values of $K_Y$. Any data points outside the bounds of the interpolation are set to the noise floor so that values outside the region of interpolation will not affect the data when the FFT is applied. This slightly reduces the SNR of the data but otherwise leaves it unaffected. Then a Hamming window is applied along the frequency dimension to remove high frequency artifacts observed in the Fourier transformed data, followed by an IFFT and an FFT shift along the frequency dimension to convert to the range domain. Then an IFFT, preceded and followed by FFT shift operations, is applied along the cross range step dimension to produce a SAR image. After the IFFTs along both dimensions are completed, the range migration is complete, and the data presented in Fig. 1(c) is converted to decibels. The SAR image's rod location was found to match that of physical measurements with a ruler, with the center of the rod being within the range and cross range resolution of 4.3 and 2 mm, respectively.

If $r$ is ever larger than the unambiguous range, well known "wrapping" or "aliasing" artifacts appear that prevent accurate reconstructions. Since frequency $f_i$ correlates with range bin $i$, this problem can be avoided in the Carrara algorithm by adding a range delay correction $\tau_D = e^{-j4\pi r_d f_i/c}$ that effectively shifts the frequency domain data a distance $r_d$ closer in $z$ so that the range becomes $r = \sqrt{(z_0 - r_d)^2 + (x - x_0)^2}$. Once this is done, the range migration algorithm perfectly interprets the range curvature produced by the scanning transceiver and accurately locates the target. To explore the sensitivity of the reconstruction on choices of range delay correction $r_d$, the imagery presented in Fig. 3 applies $r_d = 0$, 3, and 5 m to the data for each frequency in the sweep. As can be seen, the Carrara algorithm is relatively insensitive to the value of $r_d$ chosen, as long as it moves the scene of interest away from the range boundaries and toward the center. In cases where the range is roughly known, such as in the cases detailed below with scatterers embedded in a single host panel, excellent reconstructions are possible when $r_d \approx z_0$.

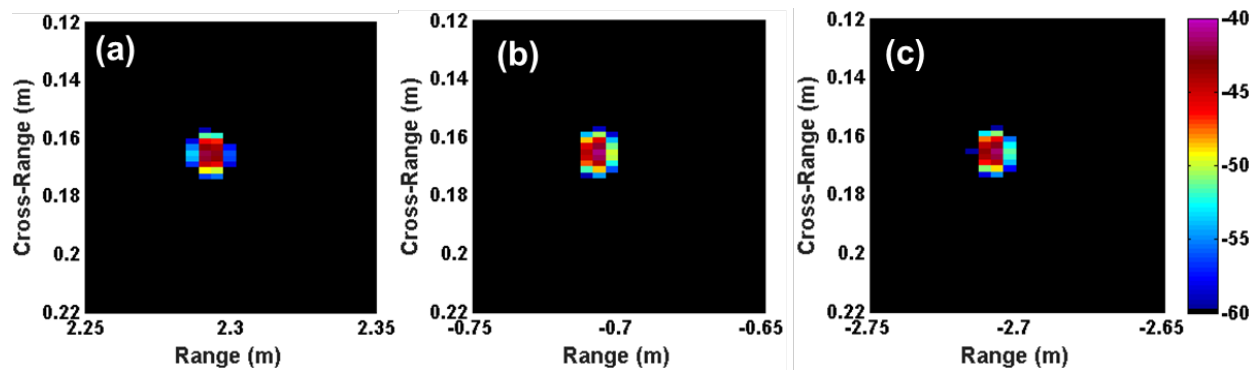

Fig. 3. SAR reconstructions plotted in dB for a single aluminum rod located 2.3 meters downrange on the Z axis and 0.17m cross-range on the X-axis with a range delay of (a) 0 m, (b) 3 m, and (c) 5 m.

**B. Foam blocks with embedded metallic scatterers**

We now apply this methodology to reconstruct more complex targets and explore the opportunities and limitations of this technique. To begin, a relatively transparent, hollow 254 mm x 305 mm x 115 mm Styrofoam packing case was used to host several different sized metallic objects so that their reconstructed locations can be compared with truth data. The Styrofoam was embedded with several staples and 10 mm long wires, and it had a small aluminum rod on top and a large aluminum rod hidden in the hollow volume behind (Fig. 4). The rail SAR was located to the left of the target, and the target was mounted upright on the base measured by the wooden ruler in Fig. 4(a). The target was located 2290 mm from the rail, and the 2D Carrara reconstructions used $r_d = 2300$ mm, so the actual range from the rail to the target is the value plotted in Fig. 4(b) plus $r_d$. The cross-range location is referenced to an arbitrarily chosen location along the rail.

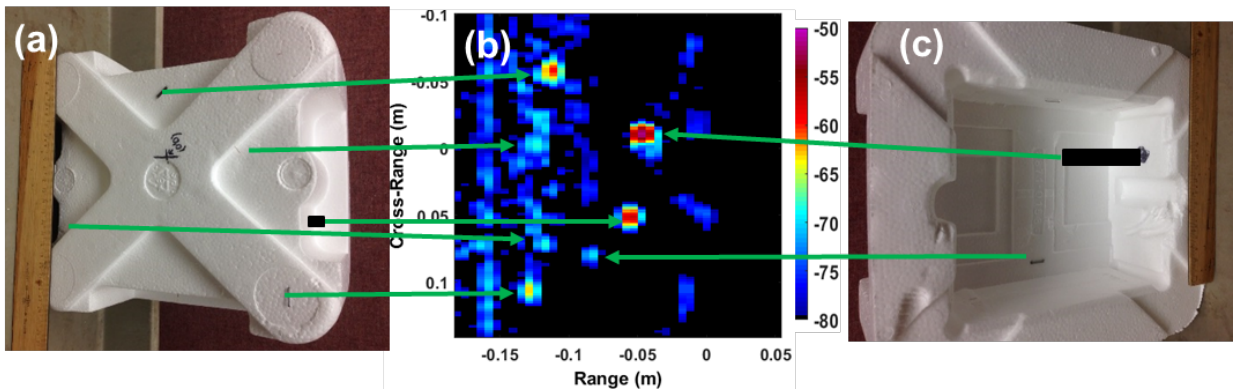

Fig. 4. (a) Photograph of the front of the Styrofoam block. (b) Reconstructed top view SAR image with strengths in dB. (c) Photograph of the back of the Styrofoam block. Objects detected by the SAR reconstruction are identified with the actual scatterers.

This variety of objects allowed us to measure the correlation of signal strength with object size. Although the small pieces of wire were difficult to discern because of their small size and orientation, the staples and rods were easily identified even though they were visibly obscured (Fig. 4). All the objects were located to within the range and cross range bin size tolerances of the locations measured with a ruler. The visibly obscured rod was 102 mm tall and 12 mm diameter and had a 40 dB SNR. The smaller, visible rod was 25mm tall and 6 mm diameter and had a 33 dB SNR. The visible staples had 27-30 dB SNR while the non-visible staple had an SNR of 20 dB. The small, visible pieces of wire had an SNR of 20 dB as well; however, the local noise levels were 10 dB higher. Of course all of the objects were found within the 115 mm depth of the host, but because it was mounted vertically, no information about they distribution of the scatterers in the Y-dimension was possible. The clutter produced by scatter from the host medium was comparatively low, and none of it produced a return with and SNR greater than 15 dB. Of course, the various methods available to reduce noise (decreased cross range step size, increased acquisition time at each step) or eliminate calibration artifacts would further improve the ability to detect objects and increase their SNR.

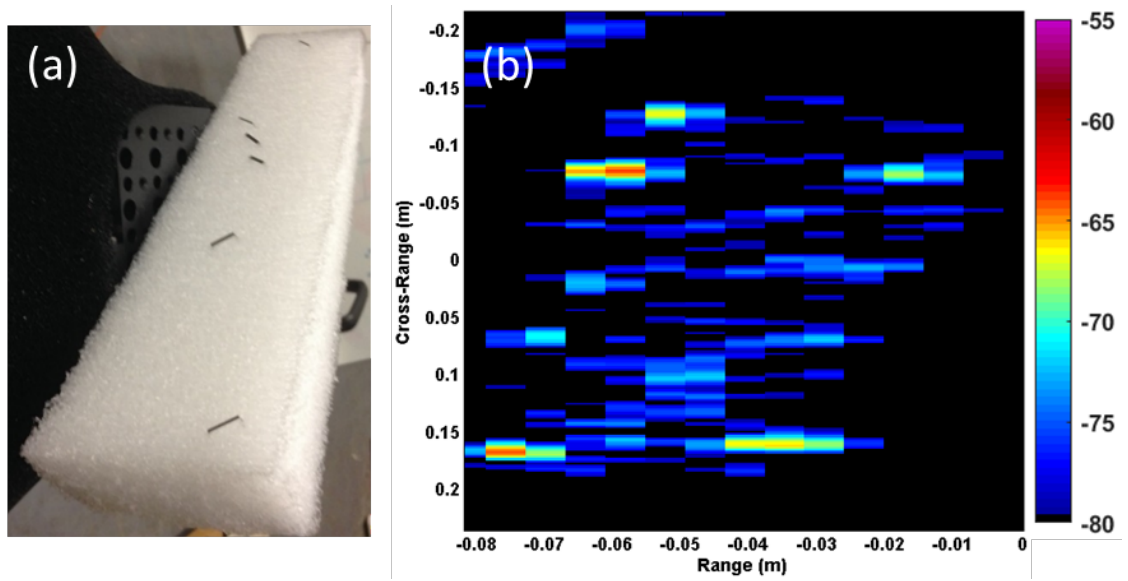

Fig. 5. (a) A photograph of the front face of the waxy packing foam embedded with staples and (b) the SAR reconstruction of the target with strengths in dB.

Next, a solid, semi-transparent, highly scattering rectangular piece of waxy packing foam 76 mm x 305 mm x 38 mm hosted six staples inserted in the front face (Fig. 5(a)) and four in the back face. The center of the target was located 2295 mm from the rail, and the 2D Carrara reconstructions used $r_d$ = 2300 mm. The foam was placed at a 30° angle rotated about the Y axis through the foam's center so that the staples were projected along the X-Z plane, removing range and location ambiguity for staples distributed along the Y direction. All the staples in the front face of the foam were located (Fig. 5(b)), but because the packing foam was less transparent and more highly scattering than the Styrofoam host with voids as large as 100 mm$^3$, many of the staples in the back were harder to discriminate from the clutter produced by the host. The clutter was 10 dB above the noise level. The staples on the front had a Signal to Clutter Ratio (SCR) between 10-15 dB, where the clutter level is determined by the scatter from the foam in regions where no metallic scatterers have been embedded. The staples on the back had an SCR 3 dB lower but were still discernable from the clutter. There was some false positives that had as much as 8 dB SCR.

### C. Tilted Celotex panels with embedded with metallic scatterers

To consider next the case of an absorbing, highly scattering host, a 305 mm x 305 mm x 13 mm section of Celotex paneling was embedded with several metallic scatterers distributed in the front and back faces of the panel. These scatterers were metallic fragments with a variety of shapes and sizes and were completely buried inside the Celotex panel as a blind test. This target approximates many practical applications for which this technique may prove useful, such as locating nails and screws in sheetrock or fragments in an arena test, and the objective here was to ascertain the penetration and resolution possible by this rail SAR technique. Although the size and location of the pre-embedded fragments could not be accurately measured, they could be approximated by the size and location of the disturbed Celotex surface region that covered them. These estimates are presented in Table 1, where the X/Z origin is the bottom left of the image and SAR reconstruction at (range, cross-range) location of (-64 mm, 155 mm). Rather than mounting the panel vertically in the X-Y plane, the panel was pivoted 47° down from vertical about the X-axis, (Fig. 6) to remove the Y ambiguity and allow the scatterers to be located by projection in the X-Z plane while maintaining strong returns from them.

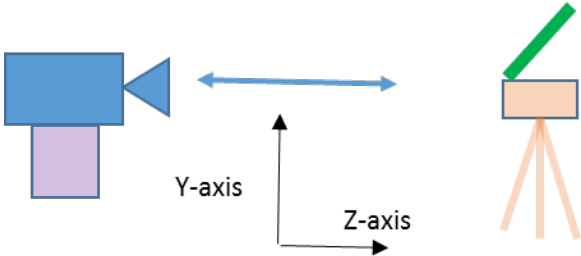

Fig. 6. Side view of the SAR setup. The range domain corresponds to the Z-axis and the transceiver moved along the rail along the X-axis into the page. The Celotex panel (green) is rotated 45° along the X-axis.

| Cross Range (mm) | Flat Range (mm) | Tilted Range (mm) | Approx. Cross Section (mm$^2$) | Front SCR (dB) | Back SCR (dB) |
|---|---|---|---|---|---|
| -115 | 280 | 141 | 80 | 6 | 10 |
| -125 | 177 | 65 | 310 | 7 | 11 |
| -30 | 225 | 101 | 50 | 4 | 4 |
| 105 | 250 | 119 | 20 | 5 | 7 |
| 80 | 227 | 102 | 7 | 6 | 4 |
| 80 | 200 | 82 | 7 | 4 | NA |
| 91 | 152 | 47 | 80 | 4 | 6 |
| 95 | 110 | 16 | 7 | NA | 3 |
| 105 | 73 | -11 | 80 | NA | 10 |
| 83 | 50 | -27 | 7 | NA | 5 |

Table 1. The 2D centroid locations and approximate sizes of objects buried in a Celotex panel, referenced to the lower left corner at (-64 mm range, 155 mm cross range) with range values uncorrected (flat) and corrected for the tilt of the panel.

The reconstruction of the scene and the identification of selected scatterers is shown in Fig. 7. The 47° tilt dramatically reduced the specular reflection from the front of the Celotex panel, but there remained a significant amount of clutter noise produced by scatter from imperfections and domains in the Celotex panel that was much stronger than in the Styrofoam or packing foam. The clutter from the panel was at least 6 dB above the noise floor and as much as 10 dB in some locations, so only scatterers with SCR > 4 dB could be confidently identified. The extra clutter noise made it difficult to identify many weak scatters, but the tilt caused the clutter noise to decrease in strength near the top of the panel because of the larger range, so weak scatterers may be more easily discriminated there.

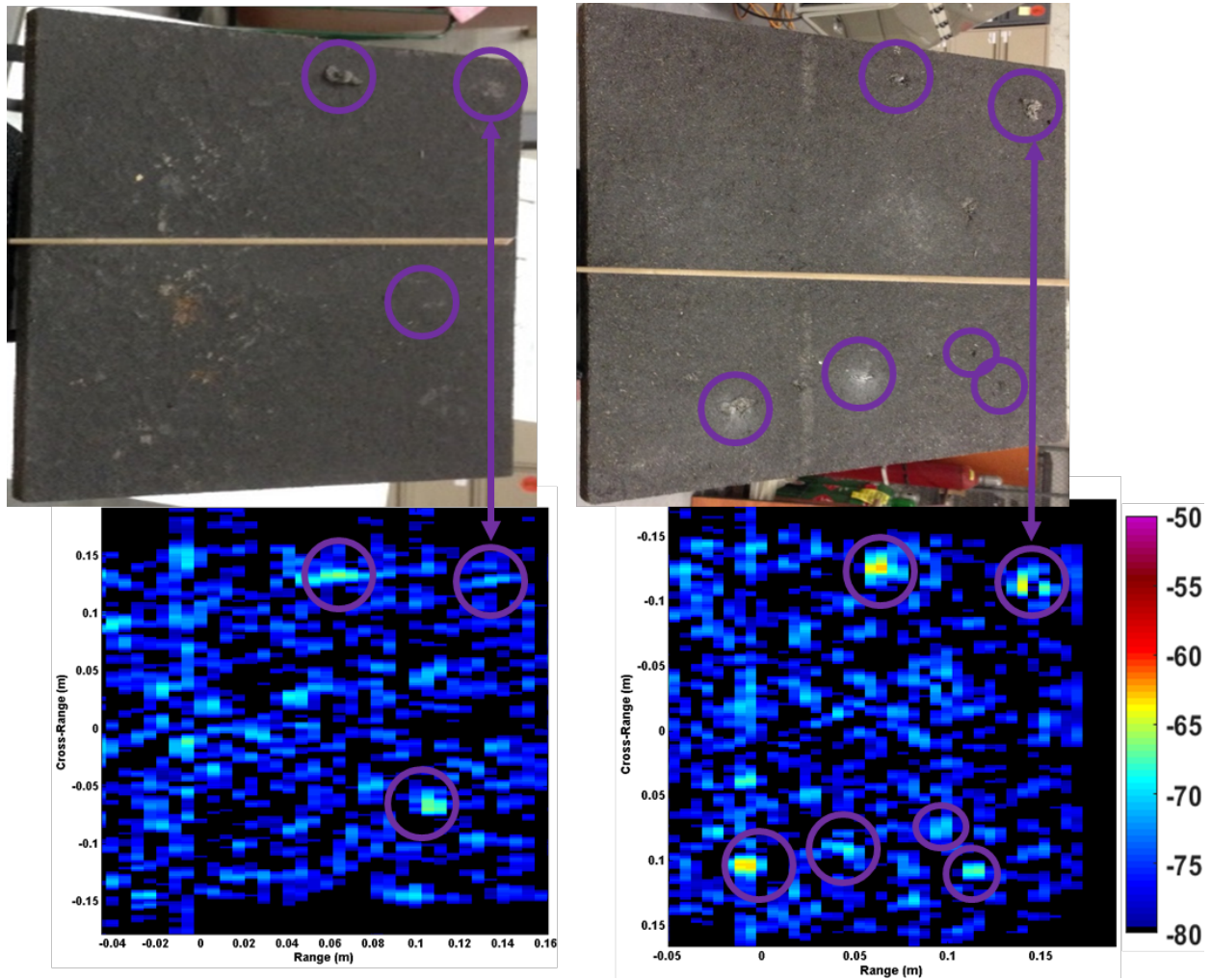

Fig. 7. The front (a) and back (b) of the Celotex panel, showing the location of the embedded metallic scatterers. The SAR reconstruction shows a correlation between objects circled in purple for the scans of the front (c) and back (d) of the Celotex panel, with strengths in dB. The two objects on the top were observed from the front and back in the SAR reconstruction.

To ascertain the ability of the rail SAR to detect and reconstruct the location of metallic scatterers in the relatively opaque, highly scattering Celotex panel, the target was measured on both sides using the same 47° tilt. The target was located 2300 mm from the rail, and the 2D Carrara reconstructions used $r_d = 2300$ mm. The SAR data confirmed what could be deduced from visual inspection of the disturbed Celotex panel surfaces: two large metallic scatterers spanned the depth of the Celotex panel and could be detected from both sides, while several smaller scatterers were only detectable from one side (Fig. 7). Regardless of orientation, the Celotex panel was so opaque that the SAR data did not confidently reveal many objects with significant SNR on the opposite side of the panel, consistent with a panel absorption coefficient of 1-10 $cm^{-1}$ at W-band.

As seen in Table 1, the two largest fragments had an SCR of 10-11 dB and could be seen on both sides of the panel, while most other scatterers could be located on one side or the other with SCR $\geq$ 4 dB. In every case, the scatterer was accurately located in range and cross-range to within the limits of the instrument's spatial resolution, recognizing that the range offset in Table 1 differs from the range measured with a ruler because of the 47° tilt of the panel, so both values are presented. Fig. 8(a) shows how the 2D locations of all the back-side scatterers, with the appropriate tilt corrections, accurately overlay the SAR 2D reconstruction of the back side of the Celotex panel. To make the scatterers easier to identify, a thresholding algorithm was applied to null range/cross-range bins of Fig. 7(d) where the SCR is below the 6 dB clutter level. Only a few small objects (<100 $mm^2$ or <-70 dBsm) on the opposite side of the panel could not be identified because their SCR was too low, and the range equation may be used to estimate the smallest size scatterer that can be detected as a function of depth in this attenuating, scattering host. Methods to increase the SCR (i.e. longer integration, more cross range steps) could be used to improve sensitivity to buried scatterers and reduce false positive detections produced by clutter.

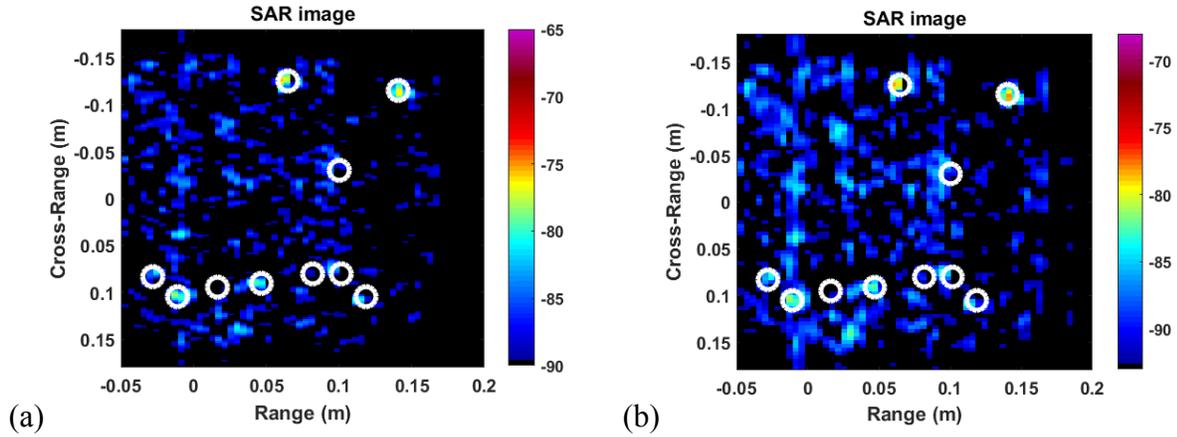

Fig. 8. The 2D centroid locations overlaid with SAR reconstruction of the tilted Celotex panel, where the threshold clutter floor values have been raised by 6 dB from Fig. 7(d). Reconstruction with (a) δx = 2 mm and (b) δx = 4 mm.

To ascertain the sensitivity to lateral step size for the transceiver, the target was imaged with cross range steps every δx = 4 mm instead of 2 mm for the same 1 m scan. Halving the number of steps would halve the acquisition time from six to three minutes but also decreased the signal to noise ratio (SNR) of the reconstruction via coherent integration, making it more difficult to identify smaller objects. The 1-2 dB degraded SCR of the returns was apparent in the data because the clutter noise from the panel was increased by a factor of two (Fig. 8(b)), thereby slightly degrading the ability to locate the scatterer. In addition to improving SCR of the strong scatterers, finer steps and more samples will allow smaller objects to be detected and distinguished from any background clutter noise, at the cost of longer acquisition times.

**D. Comparison of W-band and THz-Band using aluminum rods**

Finally, to compare the resolution of W band and THz-band rail SAR, a free space target composed of two large circular aluminum rods, a smaller circular aluminum rod, and a rotated

Page 18

rectangular aluminum post were observed at both W and THz-bands. A photograph of the target is shown in Fig. 9(a), and the locations and sizes of the objects are shown in Table 2. As mentioned above, the range resolution is five times better for the THz experiment, but the unambiguous range is also five times smaller. The cross-range resolution remains the same, as both W and THz bands used a $\delta x = 2$ mm cross range step size. The power at THz-band was significantly lower than at W-band, so the target was moved closer to the rail the range center was reduced to $r_d = 1.0$ m (THz) from $r_d = 2.3$ m (W-band) to improve SNR enough that the targets could be detected. Examining Fig. 9 thus reveals that the large cylindrical rod is located 2.34 m from the rail in the W-band experiment and 0.786 m from the rail in the THz experiment.

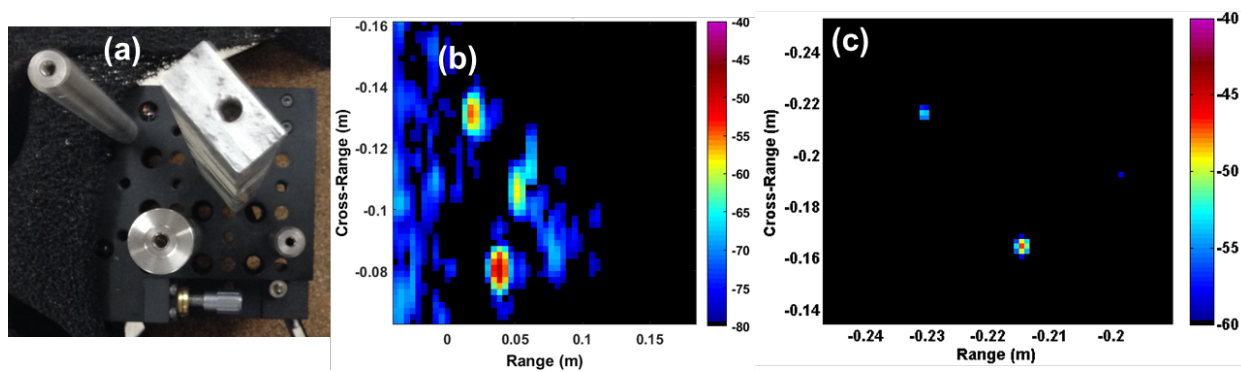

Fig. 9. (a) Photograph of the target composed of three metallic rods and a rotated rectangular rod, as viewed from above. SAR reconstructions of the target (b) at W-band and (c) the THz-band with strengths in dB.

| Range (mm) | Cross Range (mm) | Approx. Size (mm) | W band SCR (dB) | THz SCR (dB) |
|---|---|---|---|---|
| -25 | -50 | 6 D x 152 H | 35 | 37 |
| 0 | 0 | 12 D x 152 H | 40 | 43 |
| 25 | -25 | 12 L x 25 W x 305 H | 30 | 32 |
| 50 | 0 | 6 D x 102 H | 10 | 10 |

Table 2. The 2D centroid locations and approximate sizes of the aluminum posts, where D is diameter, L is length, W is width, and H is height. The relative

locations of the posts are referenced to the center of the large cylindrical rod, designated as (0,0).

Because there is no host and therefore no clutter associated with this target, the SCR and SNR are the same for each measurement, and as Table 2 indicates, strong returns were seen from all targets in both bands. It is a coincidence that the SCR for W and THz bands were within 3 dB of each other as many factors conspired to produce these strengths, including the different ranges, source power levels, and number of steps each scatterer was in the field of view. Significantly, because of the greater distance to the target in the W-band experiment afforded by the greater source power, the number of cross range steps that an object was within the field of view of the radar in W band was much greater than in the THz band, thereby increasing its SNR. Nevertheless, there is an optimal distance for target placement to maximize the competing requirements of strong illumination of the target and long duration within the field of view.

The two large aluminum rods were observed in both bands, and the spatial resolution is clearly improved in the THz band. One of the small rods was significantly obscured by the larger rod for many cross-range locations, so it didn't generate enough strong returns to achieve sufficient signal strength to be detected in the THz band, but it was weakly observed at W band. Most interestingly, the two walls of the rectangular post facing the rail were identified at W band, but these returns were quite weak. This occurred because only at a specific location on the rail did the transceiver receive a specular return from each face of the rectangular post. At all other locations of the transceiver, the signal reflected away from the transceiver, so only very weak returns from the edge were detected. The decreased range and field of view in the THz band apparently prevented this specular return from being observed, so the rectangular rod was missed. These observations suggest that spotlight SAR may be a more appropriate technique for lower power THz sources.

## 4. Conclusion

A rail SAR instrument has been constructed to operate at W-band (75 - 110 GHz) and a THz band (325 - 500 GHz), and 2D algorithms have been used to reconstruct a variety of targets composed of metallic scatterers in various hosts in order to ascertain its ability to locate small, obscured scatterers. The "top view" Carrara algorithm was able to reconstruct scenes and locate metallic scatterers at W-band with high range and cross-range resolution of 4.3 and 2 mm, respectively, while range resolution was improved to 0.86 mm at the THz band. Scatterers were easily located when embedded in target hosts of Styrofoam and packing foam that were semi-transparent at W band. Scatterers were more difficult to locate when embedded in relatively opaque Celotex panels. Only the largest scatterers that were as thick as the paneling were reliably discerned from both sides, while many smaller scatterers could only be detected from the side where they were closer to the surface. As expected, the THz band allowed for higher spatial resolution, but its lower power required the target to be located closer to the rail and caused a more limited field of view that prevented some targets from being identified. Finer cross range step sizes, longer integration times, and optimal placement to maximize power on target and duration within the field of view will improve the SNR of all targets, while more uniform host materials will improve their SCR. This work lays the foundation for a more complete 3D reconstruction of scatterers and illustrates the sensitivity of this technique to the attenuating and scattering properties of the host.